\documentclass{appolb}
\usepackage{graphicx}
\usepackage{color}
\usepackage{subfigure}
\usepackage{epsfig}
\usepackage{lipsum}
\usepackage{epstopdf}
\expandafter\let\csname equation*\endcsname\relax
\expandafter\let\csname endequation*\endcsname\relax
\usepackage{amsmath,amssymb,amsthm}    
\usepackage[numbers]{natbib}
\usepackage[titletoc
,toc,title]{appendix}
\usepackage{graphicx}
\usepackage{subfig}
\usepackage{caption}
\usepackage{sidecap}
\usepackage[applemac]{inputenc}
\usepackage{natbib}
\usepackage{graphicx}           
\usepackage{flafter}            
\usepackage{amsmath,amssymb}    
\usepackage{bm}                 
\usepackage{memhfixc}           
\usepackage[activate]{pdfcprot} 
\usepackage{pdfsync}            

\usepackage{verbatim} 

\newtheorem*{definition}{Definition}

\usepackage{algorithm,algcompatible,amsmath}
\algnewcommand\INPUT{\item[\textbf{Input:}]}%
\algnewcommand\OUTPUT{\item[\textbf{Output:}]}%

\begin{document}
\title{Understanding the dynamics of message passing algorithms: a free probability heuristics
\thanks{Presented at the conference ``Random Matrix Theory: Applications in the Information Era'' 2019 Krak\'{o}w.}%
}
\author{ Manfred Opper and Burak \c{C}akmak
\address{Department of Artificial Intelligence, Technische Universit\"{a}t Berlin, \\Berlin 10587, Germany}
}
\maketitle
\begin{abstract}
We use freeness assumptions of random matrix theory to analyze the dynamical behavior of inference algorithms for probabilistic
models with dense coupling matrices in the limit of large systems.  For a toy Ising model, we are able to recover previous results such as the property of vanishing effective memories and the analytical convergence rate of the algorithm. 
\end{abstract}
  
\bibliographystyle{plain}
\def\mathlette#1#2{{\mathchoice{\mbox{#1$\displaystyle #2$}}%
		{\mbox{#1$\textstyle #2$}}%
		{\mbox{#1$\scriptstyle #2$}}%
		{\mbox{#1$\scriptscriptstyle #2$}}}}
\newcommand{\matr}[1]{\mathlette{\boldmath}{#1}}
\newcommand{\RR}{\mathbb{R}}
\newcommand{\CC}{\mathbb{C}}
\newcommand{\NN}{\mathbb{N}}
\newcommand{\ZZ}{\mathbb{Z}}
\newcommand{\at}[2][]{#1|_{#2}}
\def\oneh{\frac{1}{2}}
\newcommand{\Kb}{{\mathbf{K}}}
\vspace{2pc}

\section{Introduction}
Probabilistic inference plays an important role in statistics, signal processing and machine learning. A major task is to compute statistics of unobserved random variables using distributions
of these variables conditioned on observed data. An exact computation of the corresponding expectations in the multivariate case is usually not possible except for simple cases. Hence, one has to resort to methods which approximate the necessary high-dimensional sums or integrals and which are often based on ideas of statistical physics \cite{mezard2009information}. A class of such approximation algorithms is often termed {\em message passing}. Prominent examples are {\em belief propagation} \cite{pearl2014probabilistic} which was developed for inference in probabilistic Bayesian networks with sparse couplings and {\em expectation propagation} (EP) which is also applicable
for networks with dense coupling matrices \cite{Minka1}. Both types of algorithms make assumptions on weak dependencies between random variables which motivate the approximation of certain expectations by Gaussian random variables invoking central limit theorem arguments \cite{Adatap}. Using ideas of the statistical physics of disordered systems, such arguments can be justified for the {\em fixed points} 
of such algorithms for large network models where couplings are drawn from random, rotation invariant matrix distributions. This extra assumption of randomness allows for further simplifications
of message passing approaches \cite{ccakmak2016self,CakmakOpper18}, leading e.g. to the {\em approximate message passing} AMP or VAMP algorithms, see \cite{Ma,rangan2019vector,takeuchi}. 

Surprisingly, random matrix assumptions also facilitate the analysis the {\em dynamical} properties of such algorithms  \cite{rangan2019vector,takeuchi,CakmakOpper19} allowing e.g. for exact computations of convergence rates \cite{CakmakOpper19,ccakmak2020analysis}. This result might not be expected, because mathematically the updates of message passing algorithms somewhat resemble the dynamical equations of spin-glass models or of recurrent neural networks which often show a complex behavior in the large system limit \cite{Mezard}. This manifests itself e.g. in a slow relaxation towards equilibrium \cite{cugliandolo1993analytical} with a possible long-time memory on initial conditions \cite{Eisfeller}. Such  properties would definitely not be ideal to the design of a numerical algorithm. 
So a natural question is: which properties of the dynamics enable both their analytical treatment and guarantee fast convergence? In this paper, we give a partial answer to this question
by interpreting recent results on the dynamics of algorithms for a toy inference problem for an Ising network. We develop a heuristics based on freeness assumptions on random matrices which lead to an understanding of the simplifications in the analytical treatment and provide a simple way for predicting the convergence rate of the algorithm.

The paper is organized as follows: In Section~2  we introduce the motivating Ising model and provide a brief presentation on the TAP mean-field equations. In Section 3 and Section~4 we present the message passing algorithm of \cite{CakmakOpper19} (to solve the TAP equations) and provide a brief discussion on its dynamical properties in the thermodynamic limit, respectively. In Section~5 and Section~6 we  recover the property of vanishing-memories and analytical convergence speed of the messaging passing algorithm using a free probability heuristic. Comparisons of our results with simulations are given in
Section 7. Section 8 presents a summary and outlook.

\section{Motivation: Ising models with random couplings and TAP mean field equations}
We consider a model of a multivariate  distribution of binary units. This is given by an Ising model with pairwise interactions of the spins $\matr s=(s_1,\ldots,s_N)^\top\in\{-1,1\}^{N}$ described by the Gibbs distribution 
\begin{equation}
p(\matr s\vert \matr J,\matr h)\doteq \frac{1}{Z}\exp\left(\frac{1}{2}\matr s^\top\matr J\matr s+\matr s^\top\matr h\right)\label{Gibbs}
\end{equation}
where $Z$ stands for the normalizing partition function.  While such models have been used for data  modeling where the couplings $\matr J$ and fields $\matr h$ are adapted to data sets \cite{hinton2007boltzmann}, we will restrict ourselves to a toy model where all external fields are equal
\begin{equation}
h_{i}=h\neq 0,~\forall i.
\end{equation}
The coupling matrix  $\matr J=\matr J^{\top}$ is assumed to be drawn at random from a rotation invariant matrix ensemble, in order to allow for nontrivial and rich classes of models. This means that $\matr J$ and $\matr V\matr J\matr V^\top$ have the same probability distributions for any orthogonal matrix $\matr V$ independent of $\matr J$. Equivalently, $\matr J$ has the spectral decomposition \cite{Collins14}
\begin{equation}
\matr J=\matr O^
\top\matr D\matr O \label{decom}
\end{equation}
where $\matr O$ is a random Haar (orthogonal) matrix that is independent of a diagonal matrix $\matr D$. This class of models generalizes the well known SK (Sherrington--Kirkpatrick) model  \cite{SK} of spin glasses for which $\matr J$ is a symmetric Gaussian random matrix.

The simplest goal of probabilistic inference would reduce to the computation of the magnetizations 
\begin{equation}
\matr {m}=\mathbb E[\matr s]
\end{equation}
where the expectation is taken over the Gibbs distribution. For random matrix ensembles, the so--called TAP equations \cite{SK} were developed in statistical physics to provide approximate solutions to $\matr {m}.$ Moreover, these equations can be assumed (under certain conditions) to give exact results (for a rigorous analysis in case of the SK model, see \cite{chatterjee2010spin}) for the magnetizations in the thermodynamic limit \cite{Mezard} $N\to\infty$ for models with random couplings. For general rotation invariant random coupling matrices, the TAP equations are given by
\begin{subequations}
	\label{tap}
	\begin{align}
	\matr m&={\rm Th}(\matr \gamma)\\
	\matr \gamma&=\matr J\matr m-{\rm R}(\chi)\matr m\\
	\chi &= \mathbb E[{\rm Th}'(\sqrt{(1-\chi){\rm R}'(\chi)} u)] \label{chi}.
	\end{align} 
\end{subequations}
Here $u$ denotes the normal Gaussian random variable and for convenience we define the function $${\rm Th}(x)\doteq\tanh(h+x) .$$ Equation (\ref{tap}) provides corrections to the simpler naive mean-field method. The latter, ignoring statistical dependencies between spins, would retain only the term $\matr J\matr m$ as the ``mean field''  acting on spin $i$. The so-called {\em Onsager reaction term}  $-{\rm R}(\chi)\matr m$ models the coherent small changes of the magnetisations of the other spins due to the presence of spin $i$. Furthermore, $\chi$ coincides with static susceptibility computed by the replica-symmetric ansatz. The Onsager term for a Gaussian matrix ensemble was developed in \cite{TAP} and later generalized to general ensembles of rotation invariant coupling matrices in \cite{Parisi} using a free energy approach. For alternative derivations, see \cite{Adatap} and \cite{CakmakOpper18}.

The only dependency on the random matrix ensemble in \eqref{tap} is via the R-transform ${\rm R}(\chi)$ and its derivative ${\rm R}'(\chi)$. The R-transform is defined as \cite{Hiai}
\begin{equation}
{\rm R}(\omega)={\rm G}^{-1}(\omega)-\frac{1}{\omega},  \label{Rtrans}
\end{equation}
where ${\rm G}^{-1}$ is the functional inverse of the Green-function 
\begin{equation}
{\rm G}(z)\doteq {\rm Tr}((z{\bf I}-\matr J)^{-1}). \label{Greens}
\end{equation}
Here, for an $N\times N$ matrix $\matr X$ we define its limiting (averaged) normalized-trace by 
\begin{equation}
 {\rm Tr}(\matr X)\doteq \lim_{N\to\infty}\frac{1}{N}\mathbb E_{\matr X}{\rm tr}(\matr X).
\end{equation}

From a practical point of view, for a concrete $N$ dimensional coupling matrix 
$\matr J$, the R-transform term can be approximated using the spectral decomposition \eqref{decom}.
The Green function \eqref{Greens} is then replaced by its empirical approximation as 
\begin{align}
{\rm G}(z)&\simeq \frac{1}{N} {\rm tr}((z{\bf I}-\matr D)^{-1}). \label{Green}
\end{align}
The R-transform ${\rm R}\doteq{\rm R}(\chi)$ (for short) and its derivative ${\rm R}'\doteq {\rm R}'(\chi)$ are then obtained by solving the fixed-point equations
\begin{subequations}
	\label{alg1}
	\begin{align}
	\lambda &={\rm R}+\frac{1}{\chi}\\
	{\rm R}&=\lambda-\frac{1}{{\rm G}(\lambda)}\\
	{\rm R}'&=\frac{1}{{\rm G}(\lambda)^2}+\frac{1}{{\rm G}'(\lambda)}\\
	\chi &= \mathbb E[{\rm Th}'(\sqrt{(1-{\rm G}(\lambda)){\rm R}'} u)].		
	\end{align} 
\end{subequations}

\section{Approximate message passing algorithm for TAP equations}
In this section we reconsider an iterative algorithm for solving the TAP equations \eqref{tap} which was introduced in \cite{CakmakOpper19} and was motivated by the so--called VAMP algorithms of \cite{rangan2019vector,takeuchi}.  We introduce a vector of auxiliary variables $\matr \gamma(t)$, where $t$ denotes the discrete time index of the iteration. We then proceed by iterating a nonlinear dynamics which is of the simple form
\begin{equation}
\label{dynamics}	
\matr \gamma(t) =\matr A f(\matr\gamma(t-1))
\end{equation} 
for $t=1,2, 3, \ldots$. Here $f$ is a nonlinear function which is applied component wise to the vector $\matr  \gamma(t-1)$ and $\matr A$ is a fixed $N\times N$ matrix. Before we specify the dynamical system (\ref{dynamics}) for the TAP equations and its parameters, we should mention that the point wise nonlinear operation followed by a matrix multiplication is typical of the dynamics of a (single layer) {\em recurrent neural network} \cite{Goodfellow-et-al-2016}. Hence, the analysis of (\ref{dynamics}) could also be of interest to these types of models. 

For the current application to the TAP equations, we specialize to the function
\begin{equation}
f(x)\doteq \frac{1}{\chi}{\rm Th}(x) -x 
\label{alg2}
\end{equation} 
where $\chi$ was defined in (\ref{chi}).
The \emph{time-independent} random matrix is given by
\begin{equation}
\matr A\doteq\frac 1\chi\left[\left(\frac 1 \chi+{\rm R}(\chi)\right){\bf I}-\matr J\right]^{-1} -\bf I.
\label{def_matrixA}
\end{equation} 
The initialization of the dynamics \eqref{dynamics} is given by $\matr \gamma(0)=\sqrt{(1-\chi){\rm R}'(\chi)} \matr u$ where $\matr u$ is a vector of independent normal Gaussian random variables.
It is easy to see that the fixed points of $\matr \gamma(t)$ coincide with the solution of the TAP equations for $\matr\gamma$, \eqref{tap}, if we identify the corresponding magnetizations by $\matr m = \chi({\matr \gamma} + f(\matr \gamma))$.

We have the following important properties of the dynamics
\begin{equation}
{\rm Tr}(\matr A)=0~~\text{and}~~ {\rm Tr}(\matr E(t))=0 ~~\text{with}~~[\matr E(t)]_{ij}\doteq f'(\gamma_i(t))\delta_{ij}, \forall t.\label{vtrace}
\end{equation}
Here, the first and second equalities follow by the constructions of the random matrix $\matr A$ and random initialization $\matr \gamma(0)$, respectively \cite{CakmakOpper19}. It is also worth mentioning that we have the freedom to replace the function $f$ with an appropriate sequence of function, say $f_t$, in such a way that the conditions ${\rm Tr}[{\rm diag}(f_t'(\matr\gamma(t)))]=0$ and $f_t\to f$ as $t\to \infty$ are fulfilled, see \cite[Section~VIII.B]{CakmakOpper19}.

\section{Dynamics in the thermodynamic limit}
Dynamical properties of fully connected disordered systems can be analyzed by a discrete time version of the dynamical functional theory (DFT) of statistical physics originally developed by Martin, Siggia and Rose \cite{Martin} and later used for the study of spin-glass dynamics, see e.g. \cite{sompo82,Eisfeller,Opper16}, and neural network models \cite{sompolinsky1988chaos}.  Using this approach, it is possible to perform the average over the random matrix ensemble of $\matr A$ and initial conditions for $N\to\infty$ and marginalize out all degrees of freedom $\gamma_j(t)$ for $j\neq i$ and all times $t$ to obtain the statistical properties of trajectories of length $T$ for an arbitrary single node $\{\gamma_i(t)\}_{t=1}^T$. Since the nodes are exchangeable random variables under the random matrix assumption, one can obtain the convergence properties of the algorithm by studying a single node.

For a rotation-invariant matrix $\matr A$ and an arbitrary function $f$, the DFT yields an ``effective'' stochastic dynamics for $\gamma_i(t)$  which is of the universal form (we skip the index $i$, since it is the same for all nodes)
\begin{equation}
\label{esp}
\gamma(t) = \sum_{s < t}{\hat {\mathcal G}(t,s)}f(\gamma(s-1)) + \phi(t), \quad t\leq T.
\end{equation}
Here $\phi(t)$ is a colored Gaussian noise term. This dynamics is of a ``mean field'' type because the statistics of the noise must be computed from averages over the process itself which involves the function $f$ and the ${\rm R}$ transform \cite{Opper16}. In general, the explicit analysis of the the single node statistics becomes complicated by the presence of the additional memory terms ${\hat {\mathcal G}(t,s)}$ which can be explicitly represented as a function of the $T\times T$ order parameter matrix 
\begin{equation}
\mathcal G(t,s)\doteq\mathbb E\left[\frac{\partial f(\gamma(t-1))}{\partial \phi (s)}\right], \quad t,s\leq T
\label{resp}
\end{equation}
which again must be computed from the entire ensemble of trajectories of  $\gamma(t)$. $\mathcal G(t,s)$ represents the average (linear) response of the variable $f(\gamma(t-1))$ to a small perturbation of the driving force $\phi(s)$ at previous times. Hence, by causality $\mathcal G$ is an upper triangular matrix (i.e. $\mathcal G(t,s)=0$ for $s \geq t$). Also, the case of zero response matrix $\mathcal G=\matr 0$ leads to ${\hat {\mathcal G}}=\matr 0$. The combination of the Gaussian noise and the response function in the dynamics has an intuitive meaning: The Gaussian can be understood as a representation of the incoherent addition of random variables arising from the multiplication of the vector $f(\matr\gamma(t-1))$ with the random matrix $\matr A$. On the other hand, by treating the typically small matrix elements $A_{ij}$ in a perturbative way \cite[Chapter~6]{Mezard}, one can estimate the influence of a node $i$ (using a linear response argument) on the $N-1$ neighboring nodes $j\neq i$, which by the symmetry of the matrix, will lead to a coherent, retarded influence of all nodes $j$ back on node $i$ at later times. This explains, why memory terms were found to be absent for neural network  dynamics with i.i.d. {\em non symmetric} random couplings \cite{sompolinsky1988chaos}. This has made a complete analytical treatment of the effective dynamics in such a case possible.

Surprisingly, for the non-linear function $f$ given in eq. \eqref{alg2} and the {\em symmetric} matrix $\matr A$, we have shown in \cite{CakmakOpper19} that the response functions (\ref{resp}) vanish, i.e. $\mathcal G(t,s) = 0$ for all $t,s$. As a result also the memory terms vanish; $\gamma (t)$ in \eqref{esp} simply becomes a Gaussian field. Hence an analytical treatment is possible as was also shown in the previous studies \cite{rangan2019vector,takeuchi}. In the following section we will use the freeness argument of random matrix theory to explain this result.

\section{Absence of memory terms and asymptotic freeness}\label{gmfc}
To analyze the average response (\ref{resp}) for a single node, we use the chain rule in the dynamical susceptibility for the original $N$ node dynamics (see \eqref{dynamics})
\begin{equation}
G_{ij}(t,s) \doteq \frac{\partial{f(\gamma_i(t-1))}}{\partial\gamma_j(s)} =
\left[(\matr E(s){\matr A}\matr E(s+1)\matr A\cdots\matr E(t-2){\matr A}{\matr E(t-1)})\right]_{ij} \label{jacob},\quad s<t.
\end{equation} 
By its construction, we can argue that the derivative w.r.t. $\gamma_i(s)$ acts in the same way as the derivative w.r.t. $\phi(s)$ and thus we will have (as $N\to \infty$)
\begin{equation}
\mathbb E\left[G_{ii}(t,s)\right] \to  \mathcal G(t,s).
\end{equation}
Here $\{G_{ii}(t,s)\}_{i\leq N}$ are random w.r.t. the random matrix $\matr A$ and random initialization $\matr\gamma(0)$. By ex-changeability $G_{ii}(t,s)\sim G_{jj}(t,s),j\neq i$, the condition ${\rm Tr}(\matr E(t))=0$ (see \eqref{vtrace}) implies vanishing single-step memories, i.e. $\mathbb E[G_{ii}(t,t-1)]\to 0$. We next argue that for further time-lags the memories do vanish \emph{in a stronger sense}. Specifically, we will show that
\begin{equation}
\epsilon(t,s)\doteq \lim_{N\to \infty}\mathbb E[G_{ii}(t,s)^2]=0,\quad s<t-1.\label{epsilon}
\end{equation}

To this end, we introduce an auxiliary random diagonal $N\times N$ matrix $\matr Z$ which is independent of $\matr A$ and $\{\matr E(t)\}$. The diagonal entries of $\matr Z$ are independent and composed of $\pm 1$ with equal probabilities. Note that ${\mathbb E}[Z_{nn}Z_{kk}]=\delta_{nk}$. Hence, we can write 
\begin{align}
\frac{1}{N}\mathbb E\left[{\rm tr}((\matr Z\matr G(t,s))^2)\right]&=\frac{1}{N}\sum_{i,j\leq N}{\mathbb E}[Z_{ii}Z_{jj}]\mathbb E[G_{ij}(t,s)G_{ji}(t,s)]\\
&=\frac 1 N\sum_{j\leq N} \mathbb E[G_{jj}(t,s)^2]=\mathbb E[G_{ii}(t,s)^2].
\end{align}
Then, we have
\begin{align}
\epsilon(t,s)&={\rm Tr} (\matr Z\matr E(s)\matr A\matr E(s+1)\cdots \matr A\matr E(t-1)\matr Z\matr E(s)\matr A\matr E(s+1)\cdots\matr A\matr E(t-1))\nonumber \\
&={\rm Tr} (\matr E_Z(t,s)\matr A\matr E(s+1)\cdots \matr A\matr E_Z(t,s)\matr A\matr E(s+1)\cdots\matr A). \label{prod}
\end{align}
Here, we have defined the diagonal matrix $\matr E_Z(t,s)\doteq\matr E(t-1)\matr Z\matr E(s)$. To simplify \eqref{prod} we will make us of the concept of \emph{asymptotic freeness} of random matrices. 
\begin{definition}\cite{Hiai}
For the two families of matrices ${\mathcal A\doteq \{\matr A_1,\matr A_2,\ldots,\matr A_a\}}$ and ${\mathcal E\doteq \{\matr E_1,\matr E_2,\ldots,\matr E_e\}}$ let ${\matr P_i(\mathcal A)}$ and ${\matr Q_i(\mathcal E)}$ stand for (non-commutative) polynomials of the matrices in ${\mathcal A}$ and the matrices in ${\mathcal E}$, respectively. Then, we say the families ${\mathcal A}$ and ${\mathcal E}$ are asymptotically free if for all  $i\in[1,K]$
and for all polynomials ${\matr P_i(\mathcal A)}$ and ${\matr Q_i(\mathcal E)}$ we have 
\begin{equation}
{\rm Tr}({\matr P_1(\mathcal A)}{\matr Q_1(\mathcal E)}{\matr P_2(\mathcal Q)}{\matr Q_2(\mathcal E)}\cdots {\matr P_K(\mathcal A)}{\matr Q_K(\mathcal E)})=0 \label{key}
\end{equation}
given that all polynomials in \eqref{key} are centered around their limiting normalized-traces, i.e.
\begin{equation*}
{\rm Tr}({\matr P_i(\mathcal A)})={\rm Tr}({\matr Q_i(\mathcal E)})=0, \quad \forall i.
\end{equation*}
\end{definition}
Namely, the limiting normalized-trace of any adjacent product of powers of matrices---which belong to different free families and are centered around their limiting normalized-traces---vanishes.	

In the product \eqref{prod} the matrices belong to two families: rotation invariant and diagonal. Under certain technical conditions---which includes the independence of matrix families---these two matrix families can be treated as asymptotically free \cite{Hiai}. E.g. $\matr A$ is asymptotically free of~$\matr Z$. Our {\bf heuristic assumption} is that $\matr A$ is also free of the diagonals $\{\matr E(t)\}$. A subtle point should be noted here: Being outcomes of the dynamical system, the diagonal matrices $\{\matr E(t)\}$ are not independent from $\matr A$. Nevertheless, since we expect that the diagonals $\matr E(t)$ have limiting spectral distributions, we consider that asymptotic freeness is a fair heuristic here.  

The result \eqref{epsilon} follows immediately from the asymptotic freeness assumption: we have that all adjacent factors in the product \eqref{prod} are polynomials belonging to the different free families and all matrices in the product are centered around their limiting normalized-traces.  

\section{Asymptotic of the local convergence}
We will analyze the convergence rate of the dynamics \eqref{dynamics} in terms of the following measure
\begin{align}
\mu_\gamma &\doteq \lim_{t\to\infty} \lim_{N\to \infty}\frac{\mathbb E\Vert \matr \gamma(t+1)-\matr \gamma(t)\Vert^2}{\mathbb E{\Vert \matr \gamma(t)-\matr \gamma(t-1)\Vert^2}}.
\end{align} 
To this end, we will assume that one starts the iterations at a point which is close enough to the fixed point of $\matr \gamma(t)$, denoted by $\matr\gamma^*$ such that a linearization  of the dynamics is justified. We conjecture (in accordance with our simulations) that the initialization does not affect the asymptotic rates. This means that we can substitute $\matr \gamma(t)$ by the following ``effective'' dynamics
\begin{equation}
\matr \gamma(t) = \matr \gamma^*+ \matr \epsilon(t) 
\end{equation}
with $\matr\epsilon(t)$ small enough to justify the linearised dynamics
\begin{align}
\matr \epsilon(t) &\simeq \matr A \matr E \matr \epsilon(t-1)=(\matr A\matr E)^t\matr \epsilon(0)~~ \text{with}~~[\matr E]_{ij}\doteq f'(\gamma_i^{*})\delta_{ij}.
\end{align}
Moreover, we consider a random initialization $\matr \epsilon(0)$ with $\mathbb E[\matr \epsilon(0)\matr \epsilon(0)^\top]=\sigma^2{\bf  I}$. Then, one can write
\begin{align}
\mu_\gamma &= \lim_{t\to\infty}\frac{{\rm Tr}[(\matr E\matr A-{\bf I})(\matr E\matr A)^t(\matr A\matr E-{\bf I})(\matr A\matr E)^t]}{{\rm Tr}[(\matr E\matr A-{\bf I})(\matr E\matr A)^{t-1}(\matr A\matr E-{\bf I})(\matr A \matr E)^{t-1}]}.
\end{align}
Similar to the response function we encounter the same product of two (asymptotic) trace free matrices. We then assume that $\matr A$ and $\matr E$ can be treated as free matrices. Doing so leads to
\begin{align}
{\rm Tr}[(\matr E\matr A)^{t\mp 1}(\matr A\matr E)^t]=0\quad \text{and}\quad {\rm Tr}[(\matr E\matr A)^t(\matr A\matr E)^t]={\rm Tr}(\matr A^2)^t{\rm Tr}(\matr E^2)^t.
\end{align}
So that we get the simple expression for the convergence rate as
\begin{align}
\mu_\gamma&={\rm Tr}(\matr A^2){\rm Tr}(\matr E^2).\label{sres}
\end{align}
This shows that when ${\rm Tr}(\matr A^2){\rm Tr}(\matr E^2)< 1$ we obtain local convergence of the algorithm
towards the fixed point. Moreover, a straightforward calculation shows that 
\begin{equation}
{\rm Tr}(\matr A^2){\rm Tr}(\matr E^2)=1-\frac{1-{{\rm Tr}(\matr E^2)}{\rm R}'(\chi)}{1-\chi^2{\rm R}'(\chi)}
\end{equation}
which exactly agrees with the result of the more complex DFT~calculation \cite{CakmakOpper19}. In the following section, we will support our heuristics by simulations on two instances of random matrices.

\section{Simulations}
In the sequel we illustrate the results of the free probability heuristics, i.e. \eqref{epsilon} and \eqref{sres}. Since we expect that these results are self-averaging in the large-system limit, our simulations are based on single instances of a large random matrix $\matr A$ and random initialization $\matr \gamma(0)$. In particular, we consider the empirical approximation of the limit \eqref{epsilon} as
\begin{equation}
\epsilon_{N}(t,s)\doteq\frac{1}{N}\sum_{i=1}^{N}G_{ii}(t,s)^2.\label{neweps}
\end{equation}

In Fig. 1(a) and 1(b), we  illustrate the vanishing memory property and the convergence rate of the dynamics \eqref{dynamics} for the SK model 
\begin{equation}
\matr J=\beta \matr G
\end{equation} 
where $G_{ij}$, $ 1 ≤ i < j ≤ N$, are i.i.d. centered Gaussian random variables with variance $1/N$. 
\begin{figure}
	\epsfig{file=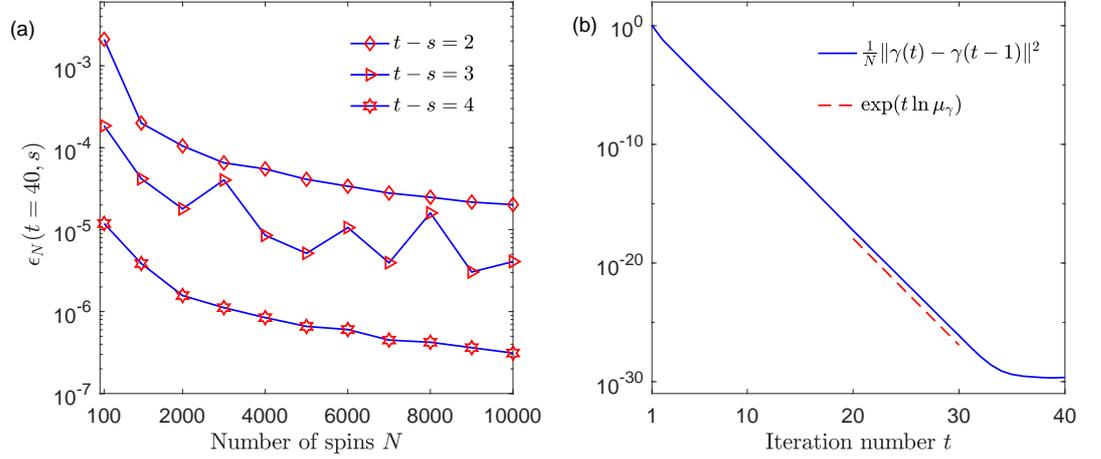,width=1.02\columnwidth}
	\vspace{.1in}
	\caption{SK model with the model parameters $h=1$ and $\beta=1$: (a) illustration of vanishing memories (w.r.t. $\epsilon_{N}(t,s)$ in \eqref{neweps}) for different time lags; (b) Asymptotic of the algorithm  with $N=10^4$ (where the flat line around $10^{-30}$ are the consequence of the machine precision of the computer which was used).}\label{fig1}
\end{figure}

Second, motivated by a recent study \cite{Greg} in random matrix theory, we consider a non-rotation invariant random coupling matrix model. The model is related to the random orthogonal model discussed by Parisi and Potters \cite{Parisi} which is defined as
\begin{equation}
\matr J=\beta\matr O^\top\matr D\matr O
\end{equation}
where $\matr O$ is a Haar matrix and $\matr D={\rm diag}(d_1,\cdots,d_N)$ has random binary elements $d_i=\mp 1$ with $\vert\{d_i=1\}\vert=N/2$. Specifically, we substitute the Haar basis of the random orthogonal model with a randomly-signed DCT (discrete-cosine-transform) matrix as
\begin{equation}
\matr J=\beta\tilde{\matr O}^\top\matr D\tilde{\matr O}~~ \text{with}~~\tilde{\matr O}\doteq \matr {\Theta}_{N}\matr Z. \label{rsh}
\end{equation}
Here, $\matr Z$ is an $N\times N$ diagonal matrix whose diagonal entries are independent and composed of binary $\mp1$ random variables with equal probabilities and $\matr \Theta$ is $N\times N$ (deterministic) DCT matrix. The simulation results for the latter model are illustrated in Figure~2. 
\begin{figure}[h]
	\epsfig{file=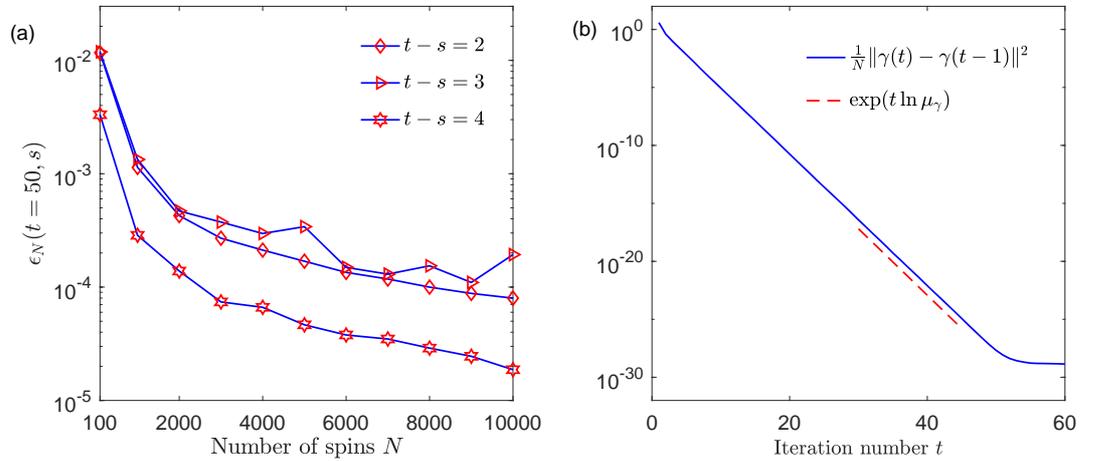,width=1.02\columnwidth}
	\vspace{.1in}
	\caption{Randomly signed DCT model with the model parameters $h=2$ and $\beta=2$: (a) illustration of vanishing memories for different time lags; (b) Asymptotic of the algorithm with $N=10^4$.}\label{fig2}
\end{figure}
They indicate that the free probability heuristics are also very accurate for randomly signed (deterministic) DCT matrix (which contains considerably less randomness compared to the rotation invariant case). As a mater of fact, this is not surprising  because for a random permutation matrix $\matr P$ and diagonal matrices $\matr D_1$ and $\matr D_2$ such that  all matrices are mutually independent,  it is proved that the  matrices $\matr P^\top\tilde{\matr O}^\top\matr D_1\tilde{\matr O}\matr P$ and $\matr D_2$ are asymptotically free \cite{Greg}. 

\section{Summary and Outlook}
In this paper we have presented a free probability heuristics for understanding and recovering 
analytical results for the dynamical behavior of so-called message passing algorithms for
probabilistic inference. Such algorithms have the form of a discrete time, recurrent neural network dynamics.  We were able to show for a toy Ising model with random couplings, that parts of previous results which were obtained by more complicated techniques can be understood and re-derived under the heuristic hypothesis of asymptotic freeness of two matrix families. Under this condition, and if matrices 
are trace free, the diagonal elements of the response function which determines the effective 
memory in the dynamics vanish. This property also yields an analytical result for the 
exponential convergence of the algorithm towards its fixed point. 
We have tested these predictions successfully on two types of  random matrix ensembles.

We expect that similar arguments can be applied to the analysis of 
more general types of inference algorithms  of the expectation propagation type. It would
also be interesting to design novel algorithms that can be analyzed assuming the freeness heuristics. Of course, the heuristics should eventually be replaced by more rigorous arguments.
While our results indicate that message passing algorithms could be analyzed
under somewhat weaker conditions on random matrices (compared to explicit assumptions on rotational invariant ensembles) the applicability of these concepts to real data needs to be shown.

\section*{Acknowledgment}
The authors would like to thank Yue M. Lu for inspiring discussions. This work was supported by the German Research Foundation, Deutsche Forschungsgemeinschaft (DFG), under Grant No. OP 45/9-1 and BMBF (German ministry of education and research) joint project 01 IS18037 A : BZML- Berlin Center for Machine Learning. 

\bibliographystyle{iopart-num}
\bibliography{mybib}

\end{document}